\documentclass[conference]{IEEEtran}
\IEEEoverridecommandlockouts
\usepackage{color}
\usepackage{multirow}
\usepackage{cite}
\usepackage{balance}
\usepackage{placeins}
\usepackage{varioref}
\usepackage{enumitem}

\usepackage{cite}
\usepackage{amsmath,amssymb,amsfonts}
\usepackage{algorithmic}
\usepackage{textcomp}
\ifCLASSINFOpdf
\usepackage[pdftex]{graphicx}
\DeclareGraphicsExtensions{.pdf,.jpeg,.png}
\else
\usepackage[dvips]{graphicx}

\DeclareGraphicsExtensions{.eps}
\fi
\graphicspath{{figs/}}
\usepackage{epstopdf}
\usepackage{booktabs}
\usepackage{amsmath}

\begin{document}
	
	\title{A Framework to Analyze Interactions between Transmission and Distribution Systems}
	

	\author{\IEEEauthorblockN{Gayathri Krishnamoorthy and Anamika Dubey}
		\IEEEauthorblockA{\textit{School of Electrical Engineering and Computer Science} \\
			\textit{Washington State University} \\
			\textit{Pullman, WA}\\
	}	
}	

\maketitle
	
	\begin{abstract}
		Analyzing the interactions between transmission and distribution (T\&D) system is becoming imperative with the increased penetrations of distributed energy resources (DERs) on electric power distribution networks. An assessment of the impacts of distribution system connected DERs on the transmission system operations is required especially pertaining to system unbalance when modeling both T\&D systems. In this paper, a coupled T\&D analysis framework is developed through co-simulation approach. The framework utilizes legacy software to separately solve the decoupled T\&D models. The T\&D interactions are captured by exchanging network solutions at the point of common coupling (PCC). The proposed co-simulation framework adopts an iterative coupling approach resulting in a co-simulation model close to a stand-alone T\&D platform. The proposed framework is tested using IEEE-9 bus transmission system model and EPRI's Ckt-24 test distribution feeder model. A case study in which the IEEE 9 bus model interfaced with three ckt24 models is also presented to demonstrate the scalability of the approach. The conditions of convergence by exchanging the boundary variables at the PCC are examined in detail using several case studies with varying levels of load unbalance.
	\end{abstract}
	
	\IEEEpeerreviewmaketitle
	
	\section{Introduction}
	
With the incentivized rapid decarbonization of electric power generation industry and aggressive renewable portfolio standards (RPS) in most states, the electric power delivery (T\&D) system is expected to transform rapidly in the foreseeable future \cite{29680},\cite{palmintier2016integrated}. Several exploratory studies and field demonstrations have pointed out that the new and recent developments including the integration of DERs, electric vehicle loads, and energy storage units are increasing the stress on power delivery systems \cite{eber2013hawaii},\cite{dubey2016analytical}.  Unfortunately, most of the existing DER interconnection studies evaluate the integration challenges of high DER penetrations either only at the distribution level or on a decoupled T\&D system \cite{evans2014new}. It is expected that the ongoing and future large-scale DER deployment projects can potentially affect the regional transmission grid operations. The situation worsens in rural areas where the distribution system is lightly loaded and covers an extended area with low load density.  The aforementioned changes are transforming the notions used to analyze the power system operations. The decoupled analysis of T\&D system is no longer adequate, calling for new tools capable of capturing the interactions between the transmission and distribution systems.

Lately, multiple frameworks to model T\&D interactions have been proposed in the literature. Based on the used approach, these frameworks are primarily categorized as following: 1) Standalone unified tools using an integrated power system modeling, 2) Co-simulation tools to combine multiple interacting domains. The “standalone” unified framework models both T\&D systems as one network \cite{evans2010verification}, \cite{chassin2014gridlab}. The major limitation of such modeling approach is the cost of simulation. Given that distribution networks contain 100s-1000s of nodes, a standalone model is usually too complex to simulate and analyze. In co-simulation approach, a hierarchical model is developed where single transmission-level representation connects to a large number of distribution systems that are run in parallel \cite{daily2014fncs}, \cite{palmintier2016final}. Unfortunately, the existing co-simulation platforms for integrated T\&D system analysis uses balanced positive sequence ac power flow for transmission system and loosely couples T\&D networks. With the increasing levels of system imbalance in the distribution system resulting from single-phase small-scale DER integration, analysis done with three-phase balanced positive sequence approach may not be sufficient to evaluate the power quality impacts. Furthermore, loosely coupling the T\&D systems limits the expandability of the existing framework to operations with faster dynamics.

A recent article addresses the above two concerns by modeling transmission system operation using sequence component analysis and iteratively coupling the T\&D systems \cite{huang2017integrated}. However, a small distribution network with 8 nodes is used in their model. The convergence properties for T\&D coupling with the increased levels of system unbalance for a large-scale distribution system require further attention. Considering the above-mentioned limitations in the existing literature, this paper presents a T\&D co-simulation framework that is close to a standalone model by accurately modeling the system unbalance and by tightly coupling the T\&D networks using an iterative approach. The following specific contributions are made in this paper.
	
\begin{itemize}[nolistsep,leftmargin=*]
	\item A three-phase transmission and distribution system power flow framework is developed that uses sequence component load flow to solve the transmission system and a three-phase unbalance power flow analysis to solve distribution system. Transmission system model is simulated using MATLAB while OpenDSS is used to model the distribution system.		
\item An iterative framework to tightly couple the T\&D networks at each iteration is developed using MATLAB. The proposed iterative method results in a co-simulation for T\&D system whose functionalities are comparable to that of the stand-alone unified models.
\item An integration of transmission system operational framework is demonstrated along with the integrated co-simulation approach by interfacing transmission economic dispatch program.
\item A detailed analysis of convergence characteristics of the proposed interactive framework is presented by simulating varying levels of load unbalance in the distribution system. Furthermore, a case study by simultaneously integrating three distribution feeders at various transmission system load nodes and associated convergence characteristics are detailed. The multiple distribution circuits are solved in parallel at each iteration of the co-simulation.
\end{itemize}


\section{Modeling the Transmission and Distribution Systems}
    The primary objective of this paper is to develop a framework that integrates T\&D systems using co-simulation. The transmission system model in MATLAB includes a detailed three-sequence network model with a 5-min ahead economic dispatch formulation solved using alternating current optimal power flow (ACOPF) model. Economic dispatch is implemented to achieve power balancing. OpenDSS, one of the commonly used distribution system modeling and analysis software, is used to simulate the three-phase unbalanced distribution system.

	\subsection{Transmission System Modeling and Analysis}
	The transmission system has been predominantly modeled as a three-phase balanced power system and solved using a positive sequence load flow analysis \cite{sun2015master}. Although this is acceptable in cases where the physical components of the transmission system are considered three-phase balanced, the positive-sequence results are inaccurate for the systems supplying unbalanced loads \cite{evans2014new}. With the proliferation of DERs in a largely unbalanced distribution system with large number of single phase customers, the positive sequence analysis is no longer adequate. A transmission system model for three-phase unbalanced analysis is developed using sequence component model. Note that the three-sequence approach to solve the transmission system is chosen over the three-phase modeling for the following reasons:
	\begin{itemize} [nolistsep,leftmargin=*]
		\item The Jacobian matrix calculation and storage is one of the most important concerns in solving a power flow problem for larger networks. Using the sequence component method effectively reduces the size of the Jacobian matrix from (6Nx6N) in three phase to (2Nx2N) for positive sequence and two (NxN) for negative and zero sequences \cite{lo1993decomposed}.
		\item The positive, negative and zero sequence components in the sequence component analysis can be solved in parallel.
		\item The computational time of the load flow problem is significantly reduced. This is desirable given the solution time required to solve and couple a large-scale distribution system.
	\end{itemize}

   The three-sequence transmission power flow operation is modeled by adopting a sequence power flow model as detailed in \cite{abdel2005improved}. For the untransposed transmission lines, the sequence admittance matrix will be full, unsymmetrical and coupled. Since the mutual coupling in sequence coupled line model is weak, it is decoupled into three independent sequence circuits by replacing the off-diagonal elements with the respective compensation current injections \cite{abdel2005improved}. The decoupled three-sequence models are solved separately. With the specified generation fixed at the beginning of the iteration for the positive sequence model, it is solved using the Newton-Raphson technique. The negative and zero sequence components of the system model are solved using linear equations \cite{abdel2005improved}.

   \subsection{Distribution System Modeling and Analysis}
   The distribution system three-phase modeling and analysis is done using OpenDSS, an open source platform designed for distribution system analysis. It performs power system studies including power flow, dynamic simulation, harmonic power flow, fault analysis etc. with simulations at various time-scales such as snapshot, daily, yearly, dynamic, and duty cycle. There is a provision in OpenDSS to include DERs at varying levels of penetrations. This allows to specify DERs' incremental capacity along with associated controls and help visualize their impacts on the distribution system. OpenDSS uses Newton's method and Fixed-point method to run three-phase distribution load flow.

    \section{Integrated T\&D Co-simulation Framework}

As detailed before, a majority of the recent studies on evaluating DER impacts on T\&D systems use a decoupled framework where the transmission system is modeled as ideal power supply for the distribution system analysis and the distribution system is modeled as lumped loads for the transmission system analysis \cite{roche2012framework},\cite{gomez2006simulating}. This decoupled framework cannot be used to visualize the exact convergence properties at the boundary of the T\&D systems. In a few recently proposed co-simulation models, T\&D systems are loosely coupled \cite{anderson2014gridspice},\cite{palmintier2016final}. These models exchange the boundary variables in the subsequent time stamps with the assumption that the changes in power system are rather slow and the system converges over multiple time steps. When intending to achieve coordinated control of T\&D systems with faster dynamics, the system boundary variables convergence should be visualized at every time-step. Fundamentally, the time stamp must not advance without making the boundary variables converge. An iteratively coupled interacting framework based on "Co-simulation" is proposed in this paper to overcome this limitation.

   \subsection{ Proposed Co-simulation Approach}
   The iterative framework based on co-simulation gives an understanding of the T\&D system operation as a whole and eliminates the uncertainties from using the decoupled model for interaction. A co-simulator is a tool that works on different set of simulators based on the time synchronization and execution coordination provided by a master algorithm. Each simulator is equipped with its own model and a solver that performs desired operations on the model \cite{palensky2017cosimulation}. The simulators are then coupled by dynamically exchanging their input and output variables with each other. The co-simulation approach stands exceptional in its operations compared to other methods of integrating T\&D systems because the individual sub-domains here are simulated within their native tools and languages. This makes their modeling easier and more coherent. That said, the transmission and distribution systems are the simulators of the co-simulation approach as shown in Fig. 1. The model within transmission simulator is the IEEE 9 bus system and three sequence load flow is used as its solver. For the distribution system simulator, EPRI Ckt 24 system is the model and the three-phase load flow from OpenDSS is used as its solver.

\vspace{-0.2cm}
    \begin{figure}[ht]
    	\centering 
    	\includegraphics[width=0.45\textwidth]{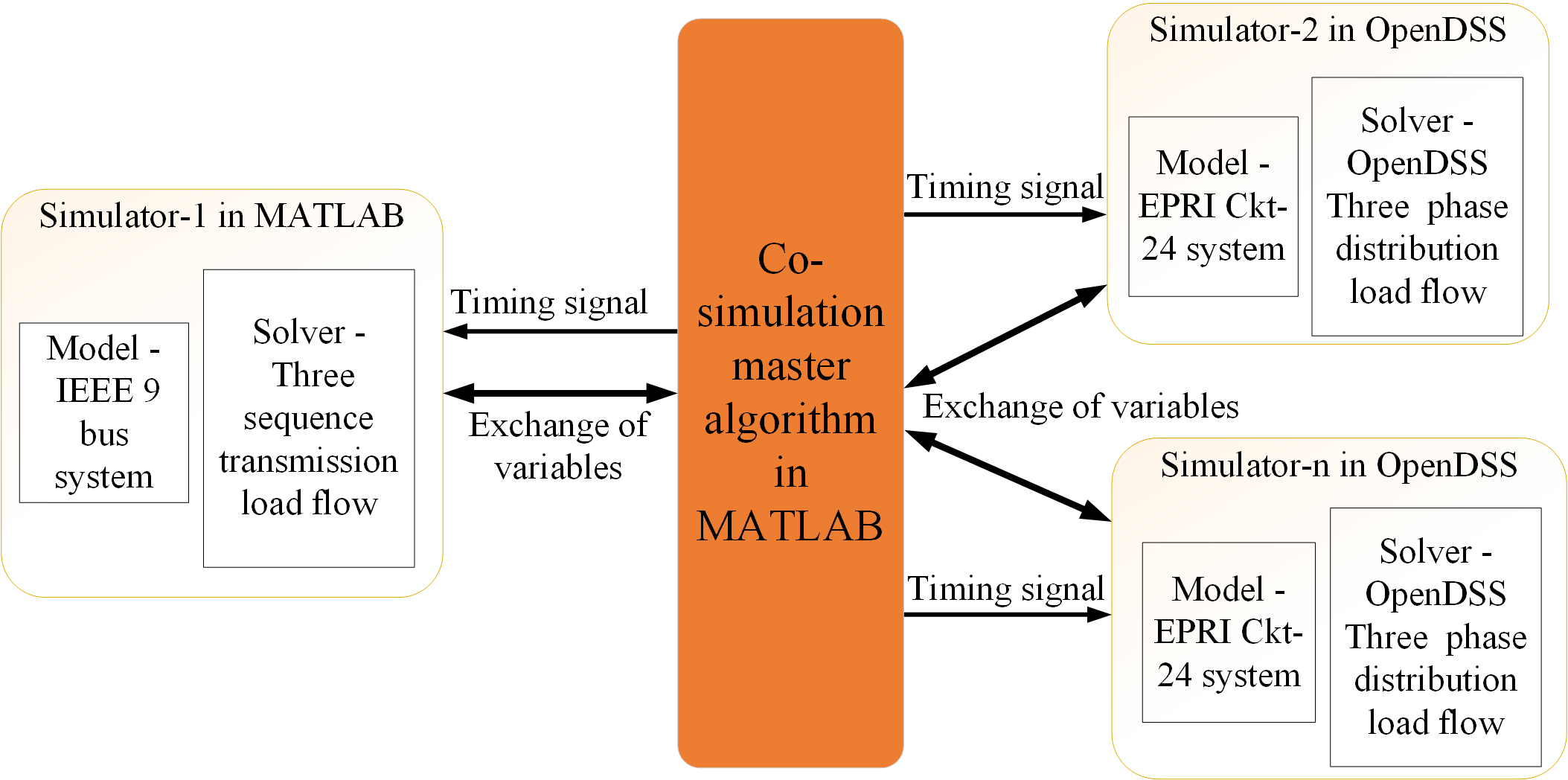}
    \caption{The Co-simulation approach to integrate T\&D systems}
    	\label{fig:1}
    	\vspace{0.1cm}
    \end{figure}

         \vspace{-0.5cm}
       \begin{figure}[ht]
      	\centering
      \includegraphics[width=0.43\textwidth]{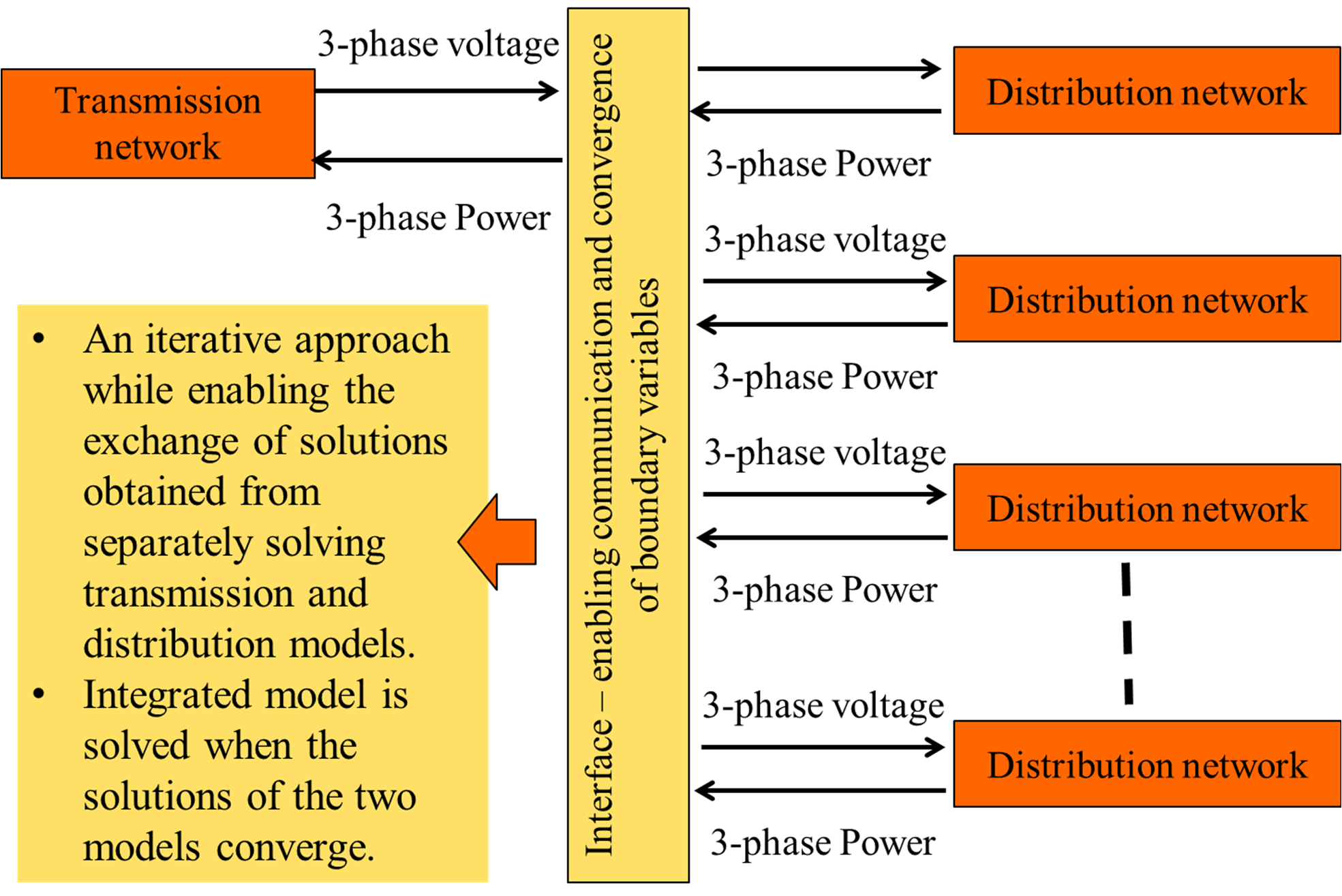}
      \vspace{-0.3cm}
      	\caption{Iterative framework for the Co-simulation approach at PCC}
      	\label{fig:2}
      	\vspace{2 pt}
      \end{figure}
\vspace{-0.3cm}

     The sequence components transmission system modeling and operations are carried out in MATLAB and the modeling of the three-phase distribution system is done using OpenDSS.  Bus voltages and angles obtained from transmission network solver and active and reactive power flow obtained from distribution network solver are interchanged at the PCC. The solutions obtained from independently solving the models are then exchanged between the two networks making the output of the transmission system an input to the distribution system and vice-versa. Exchanging the solutions follows an iterative framework as shown in Fig. 2. The time only advances after the integrated model is solved i.e. when the solutions from the individual models have converged.

    This iterative integrated framework helps us examine the convergence properties of the boundary variables exchanged between the two models. The entire co-simulation is coordinated using a master algorithm written in MATLAB. The timing components and the convergence criteria are specified in the master algorithm. Also, this co-simulation approach assists in comprehending both the subsystem level operations and the convergence at the point of common coupling (PCC). This leads to a co-simulation model that closely approximates a stand-alone unified model for the two systems.

   \subsection{T\&D Interaction Framework and   Time-Coordination}

    To describe the working of the co-simulation approach for the real-time operation of T\&D systems, the load profiles are modeled in 1-min interval for 24 hrs. The variables of exchange at the PCC are node voltages obtained from the transmission system solver and active and reactive power obtained at the PCC from the distribution system solver. Fig. 3 shows timing diagram used by the co-simulation approach to couple T\&D system operations with time-series simulation.
    \vspace{-0.6cm}
       \begin{figure}[ht]
    	\centering
    	\includegraphics[width=0.36\textwidth]{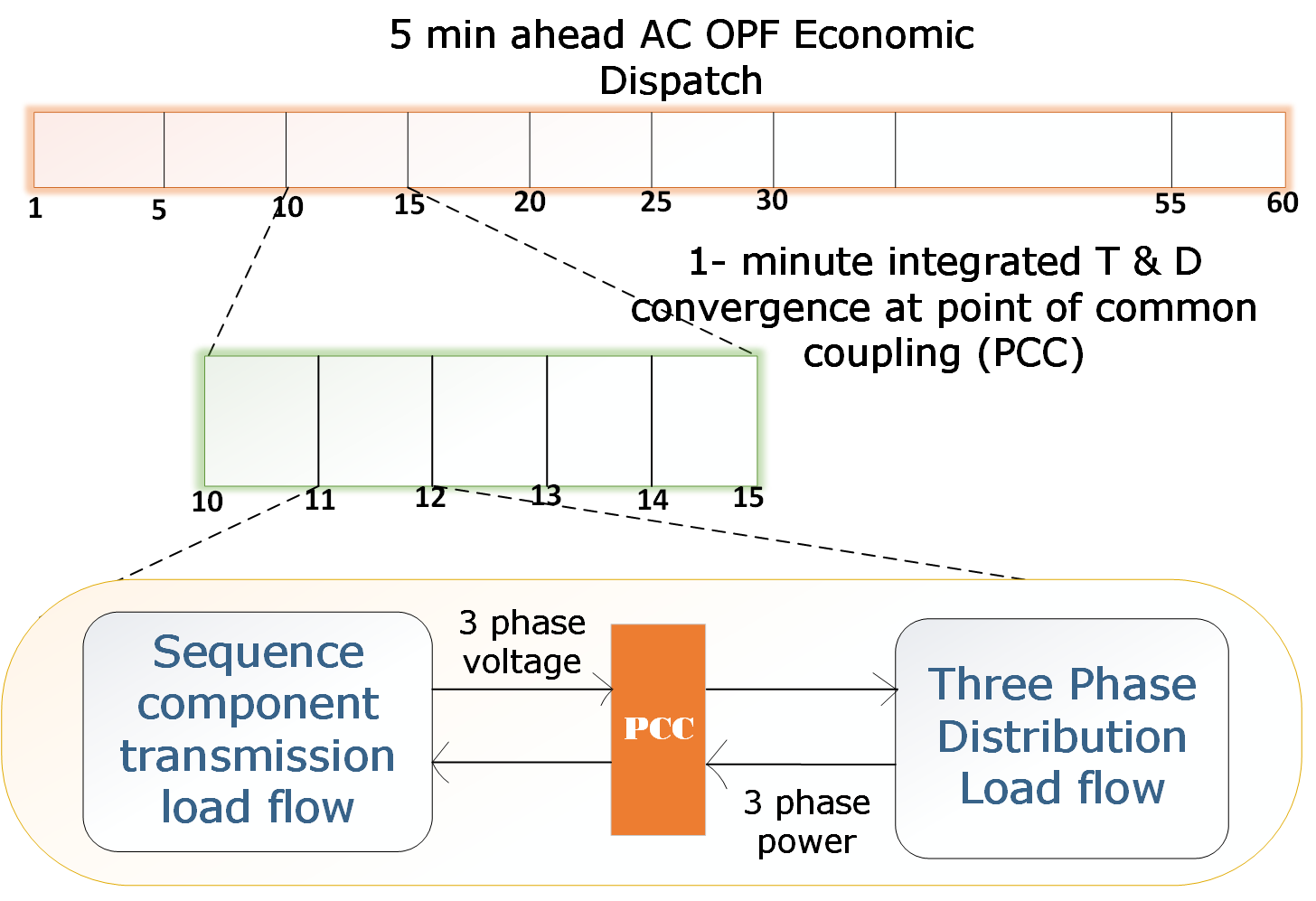}
    	\caption{Time-series simulation of the integrated T\&D system}
    	\label{fig:3}
    	\vspace{-0.2cm}
    \end{figure}

     The algorithm used for the real-time simulation of the integrated T\&D system is as follows:

    \begin{enumerate}[nolistsep,leftmargin=*]
    	\item The economic dispatch (ED) OPF formulation runs in every 5 minute discrete interval for 24 hours. At time stamp t=1 of the master algorithm, ED runs to give generator outputs based on its loading conditions at t=1.
    	\item At time stamp t=1 of the master algorithm:
    	\begin {enumerate}
    	\item The results from ED are used for solving transmission load flow. The sequence component load flow is run using MATLAB and sequence voltages are obtained.
    	\item The three sequence voltage at the bus supplying distribution feeder, termed as PCC, is converted to three-phase voltages and are provided as an input to the distribution system source node. The distribution system load flow runs based on this voltage and gives three-phase active and reactive power flow at the substation node.
    	\item The obtained three-phase active and reactive power are provided to update the transmission system model. A transmission load flow is then executed and the bus voltages at all nodes including PCC are updated.
    	\item  Step (b) is repeated until the voltage at PCC obtained at two successive iterations are within the prespecified error bound. This ensures convergence of the T\&D systems with least voltage difference at PCC.
    \end{enumerate}
        \item Once convergence is achieved at time stamp t=1, the master algorithm created in MATLAB issues a timing signal to move to the next time stamp. The transmission system uses its own load shape plot generated every one minute for 24 hrs to get the loading condition at time t=2 and initializes the load flow. The steps (a) through (d) are repeated every 1-min for next 5 intervals.
        \item After 5 time-intervals, go back to Step 1. Since the forecasted load is available in every 5-min interval, ED will be simulated every 5-min. \\
    \end{enumerate}
\vspace{-0.3cm}

  \section{Results and Discussions}

    \subsection{Simulated Case and Test Set-up}

  The transmission system used here is the IEEE 9 bus system with three generators (including the slack bus) at buses 1, 2 and 3, three loads at buses 5, 6 and 8 and three transformers connected to each of the generators. A 5-min ahead OPF formulation for economic dispatch is done in this study using the IEEE 9 bus system. The EPRI Ckt 24 is one of the real world distribution systems available in OpenDSS. This circuit has 3885 customers at 34.5 kV system voltage level with two feeders at the substation bus. The circuit total power at peak demand is 52.1MW and 11.7 MVARs. Also, Ckt 24 has a large number of single phase and two phase customers with 87 percent residential loads making the system largely unbalanced.

  Two test systems are used in this paper to check the robustness of the proposed framework. The first test system is created by replacing the cumulative load point (L6) of the transmission system as shown in Fig. 4 with Ckt24 distribution system. This test system is used as a base case scenario to test convergence at stationary load point with and without distribution load unbalances. Similarly, another test case system is created by replacing all the load points, L5, L6 and L8 of the IEEE 9-bus system by EPRI's Ckt 24 distribution circuit.
  
  \vspace{-0.4cm}
     \begin{figure}[ht]
  	\centering
  	\includegraphics[width=0.44\textwidth]{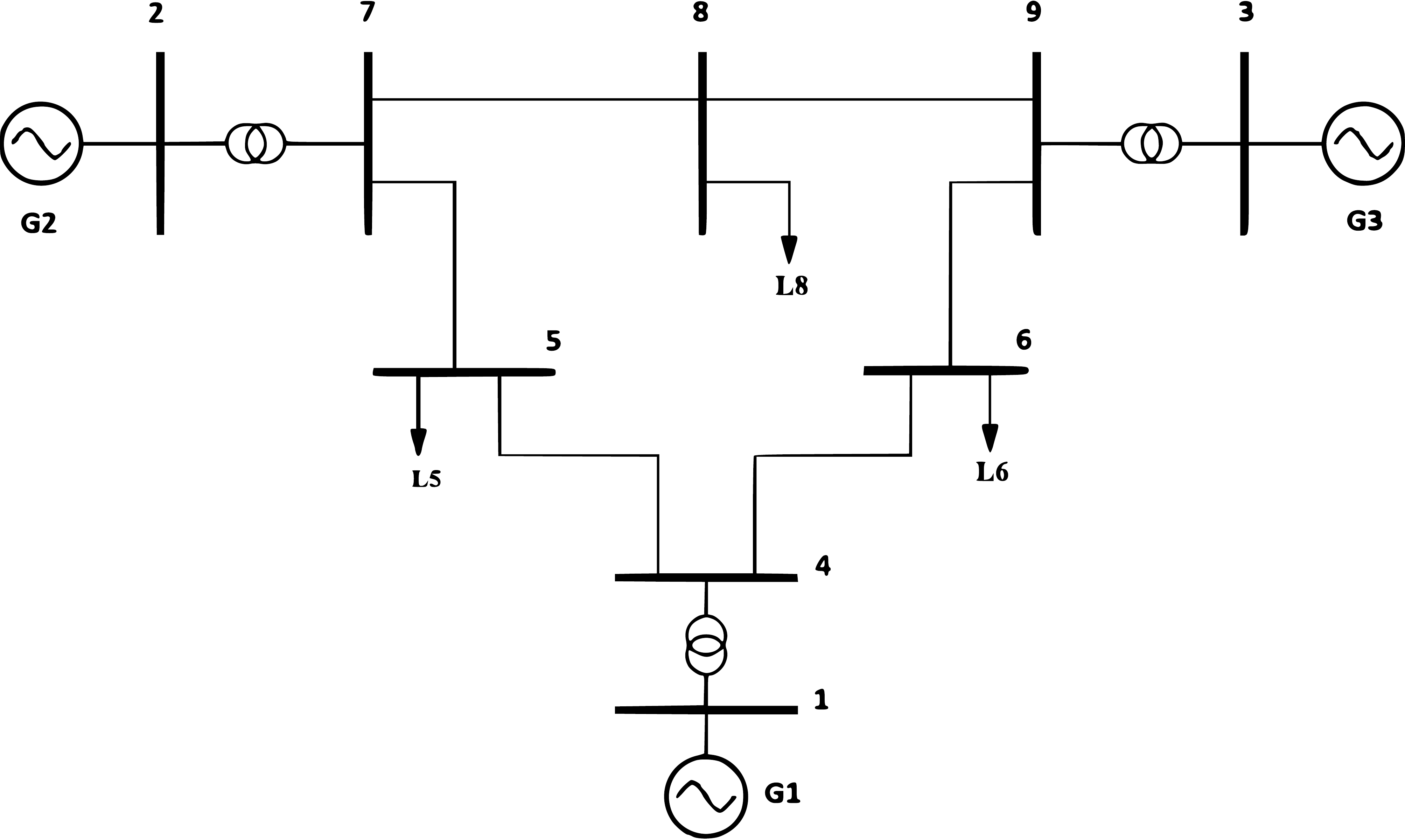}
  		\vspace{-6 pt}
  	\caption{Test System-2 with multiple distribution feeders}
  	\label{fig:4}
  	\vspace{-6 pt}
  \end{figure}

      \begin{figure}[ht]
  \includegraphics[width=0.5\textwidth]{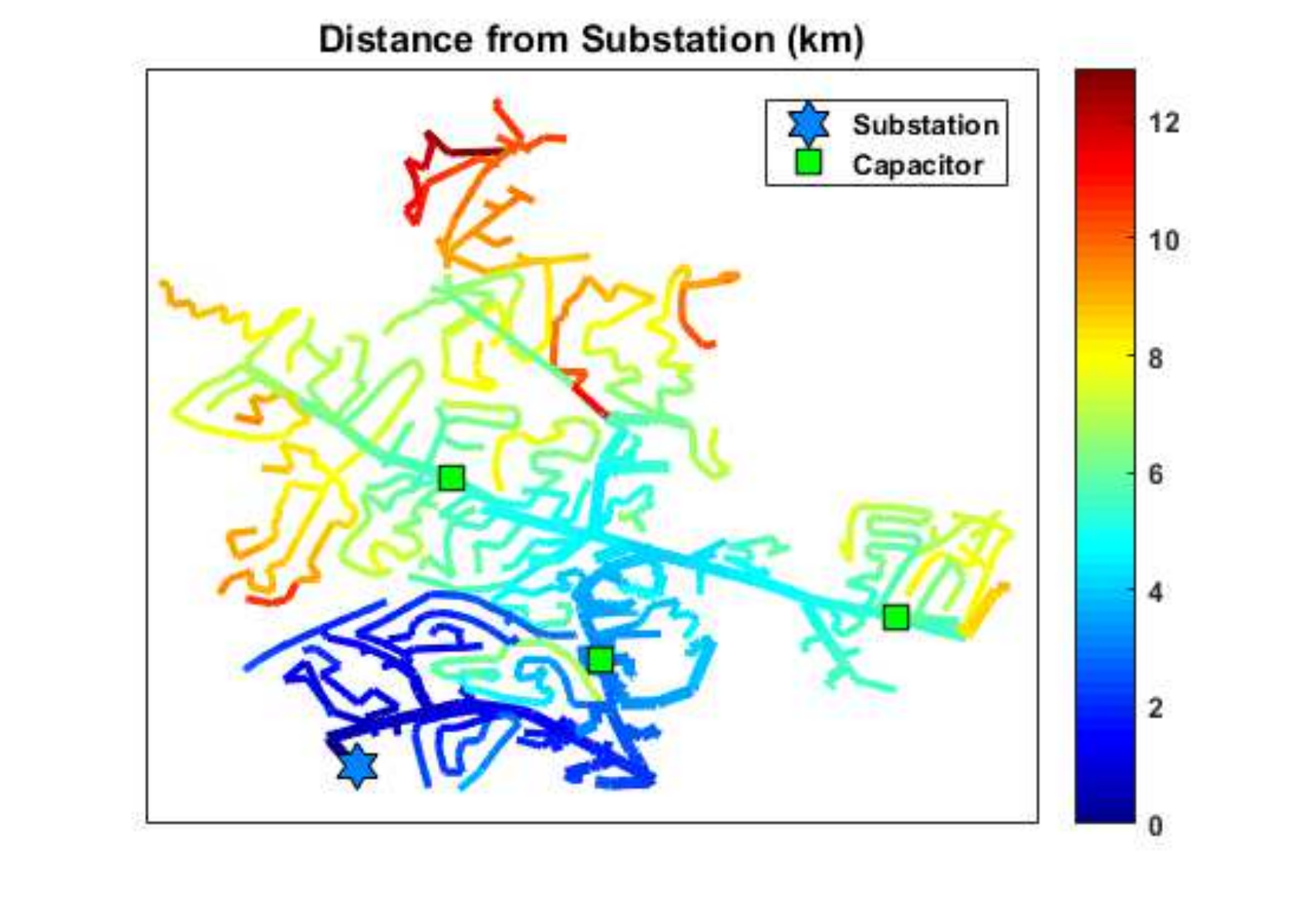}
    	\vspace{-1.2cm}
  	\caption{EPRI Ckt 24 distribution system}
  	\label{fig:5}
	\vspace{-0.2 cm}
  \end{figure}

\subsection{Snapshot T\&D Co-simulation Test}

  \subsubsection{Convergence with single distribution circuit model}
  The base case scenario here is Test system-1. This framework runs at t=1268 (in min) with the static load of 51.7 MW and 12.3 MVARs. The convergence of the system voltages over multiple iterations using the co-simulation appraoch is presented in Table 1. It can be seen from the table that phase voltages obtained from distribution and transmission system solver converge at the end of iteration 3.

\vspace{-0.3 cm}
	\begin{table}[ht]
		\centering
		\caption{Voltage convergence at T\&D PCC (bus 6) for the specified load condition}
		\label{table1}
			\setlength{\tabcolsep}{0.8em} 
				\begin{tabular} {|c|c|c|c|c|}
					\hline
					Network & Phase & Iteration 1 & Iteration 2  & Iteration 3 \\
					\hline
					\multirow{3}{4em}{IEEE 9 Bus system} & Phase A & 1.0321 & 1.0314 &  1.0312 \\
					&	Phase B	& 1.0321 & 1.0314 &  1.0312 \\
					&	Phase C	& 1.0320 & 1.0313 &  1.0311 \\
					\hline
					\multirow{3}{4em}{Ckt 24} & Phase A & 1.0299 & 1.0311 & 1.0312 \\
					&	Phase B	& 1.0300  & 1.0312 &  1.0312 \\
					&	Phase C	& 1.0299 & 1.0299 &  1.0311 \\
					\hline
				\end{tabular}
			\vspace{-5 pt}
		\end{table}

	Also, the same test case system is used to present simulation results with varying levels of distribution load unbalance. Here the percentage of unbalance in load ($\alpha$) is varied and the number of iterations taken (N) for the Test system-1 to converge is presented in Table \ref{table2}. The system converges in 4 iterations even with 15\% load unbalance. \\

	  \begin{figure*}[t]
	 	\minipage{0.32\textwidth}
	 	\includegraphics[width=\linewidth, trim={0cm 0 1cm 0},clip]{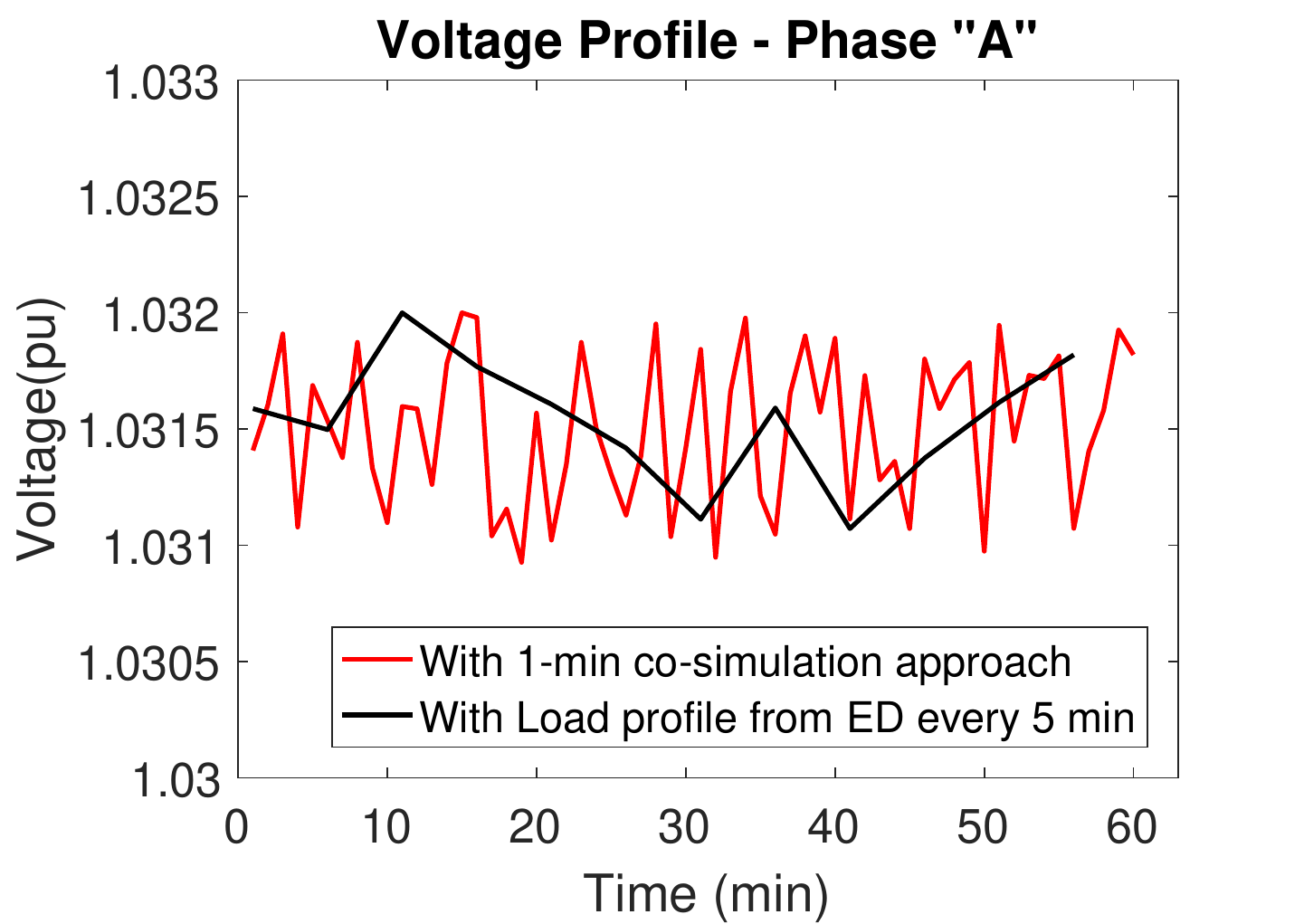}
	 	\endminipage\hfill
	 	\minipage{0.32\textwidth}
	 	\includegraphics[width=\linewidth,trim={0cm 0 1cm 0},clip]{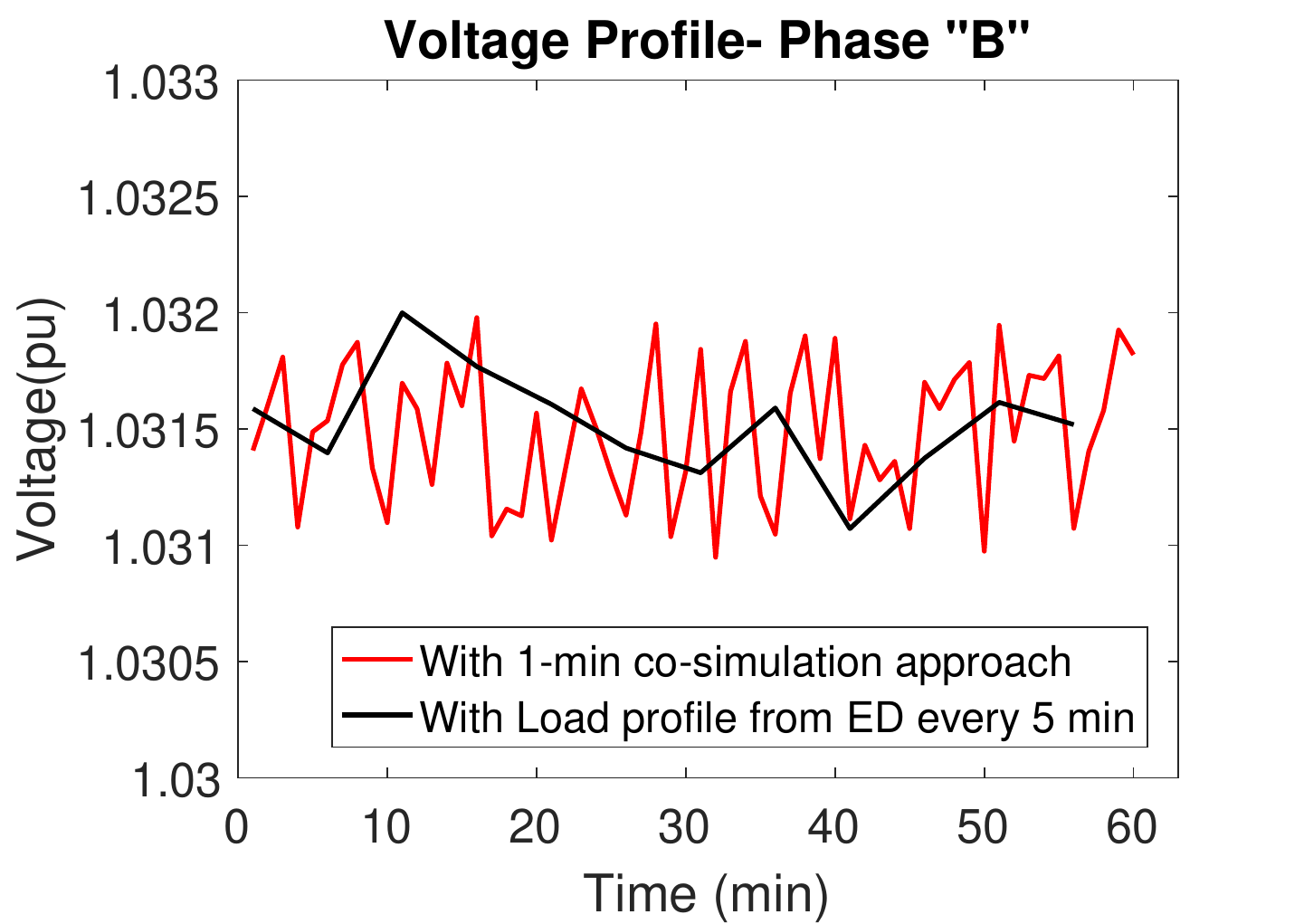}
	 	\endminipage\hfill
	 	\minipage{0.32\textwidth}%
	 	\includegraphics[width=\linewidth,trim={0cm 0 1cm 0},clip]{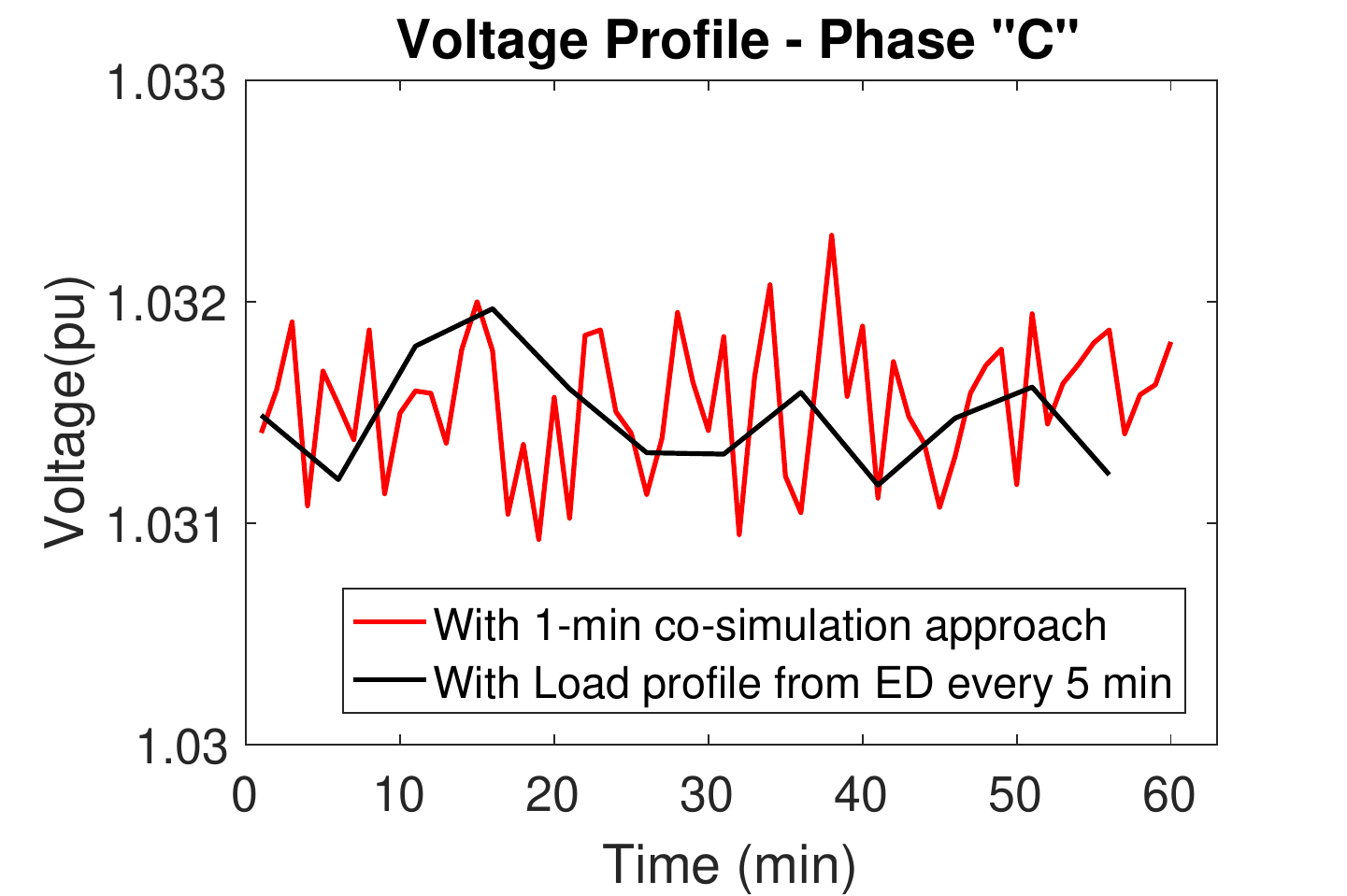}
	 	\endminipage
	 	\caption{Time-series simulation for Test system-1 with no unbalance - voltage profiles at Bus 6 for 1-hour duration}
\vspace{-0.3cm}
	 \end{figure*}
	
	\subsubsection{ Convergence with multiple distribution circuits model}
   The test case system-2 is used here for simulation where all the load points of the transmission system are replaced by the distribution system. This system is also tested for convergence with varying levels of load unbalance in the distribution circuits. The maximum  number of iterations taken for voltages to converge at individual load points and for the overall system are presented in Table \ref{table2}. For a maximum of 15\% load unbalance, the system takes no more than 8 iterations to converge.
   
   \vspace{-0.2cm}
   		\begin{table}[ht]
   	\centering
   	\caption{Number of iterations for convergence at T\&D PCC with varying levels of load unbalance}
   	\label{table2}
   	\setlength{\tabcolsep}{0.095em} 
   	\begin{tabular} {|c|c|c|c|c|c|c|}
   		\hline
   		Test & \% Unbalance & Test system-1 & \multicolumn{4}{c|}{Test system-2} \\
   		\cline{3-7}
   		cases& in loads &  N (bus 6) &  N(bus 5) & N(bus 6) & N (bus 8) & Overall N  \\
   		\hline
   		Case-1 & $\alpha=0\%$ & 3 & 3 & 3 & 3 & 3  \\
   		\hline
   		Case-2 & $\alpha=5\%$ & 3 & 3 & 5 & 5 & 5  \\
   		\hline
   		Case-3 & $\alpha=10\%$ & 4 & 4 & 6 & 6 & 6 \\
   		\hline
   	    Case-4 & $\alpha=15\%$ & 4 & 5 & 8 & 8 & 8\\
   	    \hline
   	\end{tabular}
   	\vspace{-5 pt}
   \end{table}

   \subsection{Time-series Simulation and Validation}
   The time-series simulation is carried out using co-simulation algorithm as described in Section III-B. This includes an OPF formulation for the economic dispatch giving generation and load profile for transmission system operation at every 5-min interval, and the transmission and distribution system load flow performed at every 1-min interval for 24 hours. For the time-series simulation, run every 1-min interval, the average convergence time is 7.21 sec for Test system-1 with no load unbalance and 11.78 sec for Test system-2 with load unbalance, $\alpha=15\%$. The convergence here is achieved at the same time stamp which otherwise would be performed in multiple time stamps with the decoupled system model. Also, the voltage convergence at the PCC for 1-hour simulation from t=1245-1305 (in min) is compared against the voltages obtained at the loads points on solving the decoupled transmission model. The decoupled transmission model is solved every 5-min. The three-phase voltages obtained at the PCC using the two models are compared and shown in Fig. 6. The decoupled model only approximates the trend while the coupled T\&D model obtains the actual voltage profiles.

  \section{Conclusion}
  In this paper, a real time framework for coupling T\&D system is presented. This framework includes 5 min ahead ACOPF economic dispatch formulation, three sequence transmission power flow coupled with three phase distribution power flow using a co-simulation approach. The inherent complexity of modeling the integrated T\&D systems in a single platform is addressed by iteratively coupling the models designed in their legacy softwares. This makes the framework proposed comparable to stand-alone unified models. The idea of integrating multiple distribution systems to transmission system load points is proposed and tested. This integrated T\&D framework is a valuable resource for evaluating the impacts of DER's on transmission system operations and understanding the power quality issues that would be difficult to study on a decoupled model. Also, test cases were simulated to study the integrated system convergence for varying load unbalance of the distribution system.

	\balance
	
	\bibliographystyle{ieeetr}
	\bibliography{references}

\begin{thebibliography}{10}

\bibitem{29680}
G.~L. Barbose, ``U.s. renewables portfolio standards: 2017 annual status
  report,'' 07/2017 2017.

\bibitem{palmintier2016integrated}
B.~Palmintier, E.~Hale, T.~Hansen, W.~Jones, D.~Biagioni, K.~Baker, H.~Wu,
  J.~Giraldez, H.~Sorensen, M.~Lunacek, {\em et~al.}, ``Integrated
  distribution-transmission analysis for very high penetration solar pv (final
  technical report),'' {\em National Renewable Energy Laboratory, Golden, CO,
  NREL/TP-5D00-65550}, 2016.

\bibitem{eber2013hawaii}
K.~Eber and D.~Corbus, ``Hawaii solar integration study: executive summary,''
  {\em National Renewable Energy Laboratory}, 2013.

\bibitem{dubey2016analytical}
A.~Dubey, H.~V. Padullaparti, and S.~Santoso, ``Analytical approach to estimate
  distribution circuit's energy storage accommodation capacity,'' in {\em
  Innovative Smart Grid Technologies Conference (ISGT), 2016 IEEE Power \&
  Energy Society}, pp.~1--5, IEEE, 2016.

\bibitem{evans2014new}
P.~Evans, ``New power technologies,“regional transmission and distribution
  network impacts assessment for wholesale photovoltaic generation,”
  california energy commission,'' tech. rep., CEC-200-2014-004, Aug, 2014.

\bibitem{evans2010verification}
P.~Evans, ``Verification of energynet{\textregistered} methodology,'' {\em
  California Energy Commission, CEC-500-2010-021, Dec}, 2010.

\bibitem{chassin2014gridlab}
D.~P. Chassin, J.~C. Fuller, and N.~Djilali, ``Gridlab-d: An agent-based
  simulation framework for smart grids,'' {\em Journal of Applied Mathematics},
  vol.~2014, 2014.

\bibitem{daily2014fncs}
J.~Daily, J.~Fuller, S.~Ciraci, A.~Fisher, L.~Marinovic, and K.~Agarwal,
  ``Fncs: Framework for network co-simulation,'' {\em Richland, WA}, 2014.

\bibitem{palmintier2016final}
B.~Palmintier, E.~Hale, T.~M. Hansen, W.~Jones, D.~Biagioni, K.~Baker, H.~Wu,
  J.~Giraldez, H.~Sorensen, M.~Lunacek, {\em et~al.}, ``Final technical report:
  Integrated distribution-transmission analysis for very high penetration solar
  pv,'' tech. rep., NREL (National Renewable Energy Laboratory (NREL), Golden,
  CO (United States)), 2016.

\bibitem{huang2017integrated}
Q.~Huang and V.~Vittal, ``Integrated transmission and distribution system power
  flow and dynamic simulation using mixed three-sequence/three-phase
  modeling,'' {\em IEEE Transactions on Power Systems}, vol.~32, no.~5,
  pp.~3704--3714, 2017.

\bibitem{sun2015master}
H.~Sun, Q.~Guo, B.~Zhang, Y.~Guo, Z.~Li, and J.~Wang, ``Master--slave-splitting
  based distributed global power flow method for integrated transmission and
  distribution analysis,'' {\em IEEE Transactions on Smart Grid}, vol.~6,
  no.~3, pp.~1484--1492, 2015.

\bibitem{lo1993decomposed}
K.~L. Lo and C.~Zhang, ``Decomposed three-phase power flow solution using the
  sequence component frame,'' 1993.

\bibitem{abdel2005improved}
M.~Abdel-Akher, K.~M. Nor, and A.~A. Rashid, ``Improved three-phase power-flow
  methods using sequence components,'' {\em IEEE Transactions on power
  systems}, vol.~20, no.~3, pp.~1389--1397, 2005.

\bibitem{roche2012framework}
R.~Roche, S.~Natarajan, A.~Bhattacharyya, and S.~Suryanarayanan, ``A framework
  for co-simulation of ai tools with power systems analysis software,'' in {\em
  Database and Expert Systems Applications (DEXA), 2012 23rd International
  Workshop on}, pp.~350--354, IEEE, 2012.

\bibitem{gomez2006simulating}
J.~G. G{\'o}mez-Gualdr{\'o}n and M.~Velez-Reyes, ``Simulating a multi-agent
  based self-reconfigurable electric power distribution system,'' in {\em
  Computers in Power Electronics, 2006. COMPEL'06. IEEE Workshops on},
  pp.~1--7, IEEE, 2006.

\bibitem{anderson2014gridspice}
K.~Anderson, J.~Du, A.~Narayan, and A.~El~Gamal, ``Gridspice: A distributed
  simulation platform for the smart grid,'' {\em IEEE Transactions on
  Industrial Informatics}, vol.~10, no.~4, pp.~2354--2363, 2014.

\bibitem{palensky2017cosimulation}
P.~Palensky, A.~A. Van~der Meer, C.~D. Lopez, A.~Joseph, and K.~Pan,
  ``Cosimulation of intelligent power systems: Fundamentals, software
  architecture, numerics, and coupling,'' {\em IEEE Industrial Electronics
  Magazine}, vol.~11, no.~1, pp.~34--50, 2017.

\end{thebibliography}
	
\end{document}